\documentclass[aps,pra,superscriptaddress,nofootinbib,notitlepage,showpacs,floatfix,twocolumn]{revtex4-1}
\usepackage{epsfig}
\usepackage{bm}
\usepackage{amsmath}
\usepackage{subfigure}

\newcommand{\mc}{\mathcal}

\usepackage{hyperref}
\hypersetup{
    colorlinks=true,       % false: boxed links; true: colored links
    linkcolor=cyan,          % color of internal links
    citecolor=magenta,        % color of links to bibliography
    filecolor=magenta,      % color of file links
    urlcolor=cyan,           % color of external links
    runcolor=cyan
}
\usepackage[pdftex]{color}

\begin{document}

\title{Defining and detecting  quantum speedup}

\author{Troels F. R{\o}nnow}
\affiliation{Theoretische Physik, ETH Zurich, 8093 Zurich, Switzerland}
\author{Zhihui Wang}
\affiliation{Department of Chemistry and Center for Quantum Information Science \& Technology,  University of Southern California, Los Angeles, California 90089, USA}
\author{Joshua Job}
\affiliation{Department of Physics and Center for Quantum Information Science \& Technology,  University of Southern California, Los Angeles, California 90089, USA}
\author{Sergio Boixo}
\affiliation{Google, 150 Main St, Venice Beach, CA, 90291}
\author{Sergei V. Isakov}
\affiliation{Google, Brandschenkestrasse 110, 8002 Zurich, Switzerland}
\author{David Wecker}
\affiliation{Quantum Architectures and Computation Group, Microsoft Research, Redmond, WA 98052, USA}
\author{John M. Martinis}
\affiliation{Department of Physics, University of California, Santa Barbara, CA 93106-9530, USA}
\author{Daniel A. Lidar}
\affiliation{
Departments of Electrical Engineering, Chemistry and Physics, and Center for Quantum Information Science \& Technology, University of Southern California, Los Angeles, California 90089, USA
}
\author{Matthias Troyer$^*$}
\affiliation{Theoretische Physik, ETH Zurich, 8093 Zurich, Switzerland}

\maketitle

%%%%%%%%%
\onecolumngrid
\textbf{
The development of small-scale digital and analog quantum devices raises the question of how to fairly assess and compare the computational power of classical and quantum devices, and of how to detect quantum speedup. Here we show how to define and measure quantum speedup in various scenarios, and how to avoid pitfalls that might mask or fake quantum speedup. We illustrate our discussion with data from a randomized benchmark test on a D-Wave Two device with up to 503 qubits. 
Comparing the performance of the device on  random spin glass instances with limited precision to simulated classical and quantum annealers, we find no evidence of quantum speedup when the entire data set is considered, and obtain inconclusive results when comparing subsets of instances on an instance-by-instance basis. Our results for one particular benchmark do not rule out the possibility of speedup for other classes of problems and illustrate that quantum speedup is elusive and can depend on the question posed.
}\\\\

\twocolumngrid
%\input{abstract11}
%%%%%%%%%

\section{Introduction}

The interest in quantum computing originates in the potential of a quantum computer to solve certain computational problems much faster than is possible classically. Examples are the factoring of integers \cite{Shor:94} or the simulation of quantum systems \cite{Feynman1982}. Shor's algorithm can find the prime factors of an integer in a time that scales polynomially 
in the number of digits of the integer to be factored, while all known classical algorithms scale exponentially. The simulation of the time evolution of a quantum system on a classical computer also takes exponential resources,  because the Hilbert space of an $N$ particle system is exponentially large in $N$, 
while quantum hardware can simulate the same time evolution with polynomial complexity \cite{Lloyd:1996fk,Berry:2013uq}.

In these examples the quantum algorithm is exponentially faster than the best available classical algorithm. This type of exponential quantum speedup substantially simplifies the discussion, since it renders the details of the classical or quantum hardware unimportant. According to the extended Church-Turing thesis all classical computers are equivalent up to polynomial factors \cite{Parberry:1986uo}. Similarly, all proposed models of quantum computation are polynomially equivalent, so that a finding of exponential quantum speedup will be model-independent. 
In other cases, in particular on small devices, or when the quantum speedup is polynomial, defining and detecting quantum speedup becomes more subtle. One such subtlety is how to properly define the hardness of a problem given prior knowledge about the answer \cite{smolin2013}.

Here we discuss how to define ``quantum speedup'' and show that this term may refer to different quantities depending on the goal of the study. In particular, we define what we call ``limited quantum speedup"---essentially a speedup relative to a given, corresponding classical algorithm---and explain how such a speedup can be reliably detected. To illustrate these issues we compare the performance of a $503$-qubit D-Wave Two (DW2) device to classical algorithms running on a standard CPU and analyze the evidence for quantum speedup on random spin glass problems. This example is particularly relevant since it is an open question whether quantum annealing \cite{Kadowaki1998} or the quantum adiabatic algorithm \cite{farhi} can exhibit a quantum speedup for such problems. Random spin glass problems are an interesting benchmark problem, though not necessarily the most relevant for practical applications, such as machine learning.
We also discuss issues that might mask or fake a quantum speedup when not considered carefully, such as comparing suboptimal algorithms or improperly accounting for the scaling of hardware resources. 

\section{Defining quantum speedup}

\subsection{The classical to quantum scaling ratio}

When the time to solution depends not only on the problem size $N$ but also on the specific problem instance, then the purpose of the comparison becomes another factor in deciding how to measure performance. Specifically, when a device is used as a tool for solving problems, then the question of interest is to determine which device is better for the hardest problem, or for \emph{almost all} possible problem instances. On the other hand, if we are interested in aspects of the underlying physics of a device then it might suffice to find \emph{some} instances or a \emph{subclass} of instances where a quantum device exhibits a speedup. These two questions will lead to different quantities of interest. 

In all of these cases we denote the time used by a classical device to solve a problem of size $N$ by $C(N)$ and the time used on the quantum device by $Q(N)$, defining quantum speedup as the ratio
\begin{equation}
S(N)=\frac{C(N)}{Q(N)}.
\end{equation}
Note that in both the quantum and classical case this definition includes a specific choice of algorithm and device.

The first question that arises is which classical algorithm to compare against, i.e., what is $C(N)$. This leads to different definitions of quantum speedup. 

\subsection{Five different types of quantum speedup} 

The optimal scenario is one of a \emph{provable quantum speedup}, where there exists a proof that no classical algorithm can outperform a given quantum algorithm. Perhaps the best known example is Grover's search algorithm \cite{Grover:97a}, which exhibits a provable quadratic speedup over the best possible classical algorithm \cite{Bennett:1997lh}, assuming an oracle. 

A \emph{strong quantum speedup} was defined in \cite{Traub2013} by using the performance of the \emph{best} classical algorithm for $C(N)$, whether such an algorithm is known or not. Unfortunately the performance of the best classical algorithm is unknown for many interesting problems. In the case of factoring, for example, all known classical algorithms have super-polynomial cost in the number of digits $N$ of the number to be factored \cite{pomerance}, while the cost of Shor's algorithm is polynomial in $N$. However, a proof of a classical exponential lower-bound for factorization is not known \cite{P-NP-comment}. 
 A less ambitious goal is therefore desirable, and thus one usually defines \emph{quantum speedup} (without additional adjectives) by comparing to the \emph{best available} classical algorithm instead of the best possible classical algorithm.  

However, this notion of quantum speedup depends on there being a consensus about ``best available", and this consensus may be time- and community-dependent \cite{community-comment}.
 In the absence of a consensus about what is the best classical algorithm, we define {\em potential (quantum) speedup} as a speedup compared to a specific classical algorithm or a set of classical algorithms. An example is the simulation of the time evolution of a quantum system, where the propagation of the wave function on a quantum computer would be exponentially faster than a direct integration of  Schr\"odinger's equation on a classical computer. A potential quantum speedup can of course be trivially attained by deliberately choosing a poor classical algorithm (for example, factoring using classical instead of quantum period finding while ignoring known, better classical factoring algorithms), so that here too one must make a genuine attempt to compare against the best classical algorithms known, and any potential quantum speedup might be short-lived if a better classical algorithm is found. 

Underlying all the above notions of quantum speedup is the availability of a fully coherent, universal quantum computer.
 A weaker scenario is one where the device is merely a putative or candidate quantum information processor, or where a quantum algorithm is designed to make use of quantum effects but it is not known whether these quantum effects provide an advantage over classical algorithms.  To capture this scenario, which is of central interest to us in this work, we define \emph{limited quantum speedup} as a speedup obtained when comparing specifically with classical algorithms that  ``correspond'' to the quantum algorithm in the sense that they implement the same algorithmic approach, but on classical hardware. In the context of an analog quantum device this can be thought of as being the result of decohering the device. Since there is no unique way to decohere a quantum device, one may arrive at different corresponding classical algorithms. A natural example is quantum annealing implemented on a candidate physical quantum information processor \textit{vs} either classical simulated annealing, classical spin dynamics, or simulated quantum annealing (as defined in Methods). In this comparison a limited quantum speedup would be a demonstration that quantum effects improve the annealing algorithm \cite{decohered-comment}.

\section{Classical and quantum annealing of a spin glass}

As our primary example we will use the problem of 
finding the ground state of an Ising spin glass model described by a  ``problem Hamiltonian"  
\begin{equation}
H_{\mathrm{Ising}} = -\sum_{i\in \mc{V}} h_i \sigma_i^z - \sum_{(i,j)\in \mc{E}} J_{ij} \sigma_i^z \sigma_j^z \ ,
\label{eq:H}
\end{equation}
with $N$ binary variables $\sigma_i^z = \pm 1$. The local fields $\{h_i\}$ and coupling $\{J_{ij}\}$ are fixed and define a \emph{problem instance} of the Ising model.
 The spins occupy the vertices $\mc{V}$ of a graph $G=\{\mc{V},\mc{E}\}$ with edge set $\mc{E}$. 
Solving this problem problem is NP-hard already for planar graphs~\cite{Barahona1982}, which means that no polynomial time algorithm to find these ground states is known and the computational effort of all existing classical algorithms scales exponentially with problem size. NP-hardness refers only to the hardest problems, but the typical problem in our benchmarks, where the graph forms a two-dimensional (2D) lattice, is still hard since for zero local fields ($h_i=0$) there exists a spin glass phase at zero temperature. While the critical temperature $T_c=0$ for these 2D spin glasses makes the problem easier than 3D spin glasses with a nonzero $T_c>0$ \cite{katzgraber}, solving the typical problem instance is nevertheless non-trivial and with all known algorithms a super-polynomial scaling is observed. While quantum mechanics is not expected to reduce this scaling to polynomial, a quantum algorithm might still scale better with problem size $N$ than any classical algorithm. 

We use simulated annealing (SA)  \cite{Kirkpatrick1983}, simulated quantum annealing (SQA) \cite{sqa1,Santoro}, and a DW2 device to find the ground states of the Ising model above (see Methods for details). The D-Wave devices \cite{Harris2010,0953-2048-23-6-065004,berkley2010scalable,Johnson2011} are designed to be physical realizations of quantum annealing using superconducting flux qubits and programmable couplers.
Tests on a $108$-qubit D-Wave One (DW1) device \cite{ourpaper} have shown that despite decoherence and coupling to a thermal bath, the device correlates well with SQA, which is consistent with it actually performing quantum annealing \cite{Smolin,comment-SS}. It also correlates well with the predictions of a quantum master equation \cite{Boixo2012}, which is consistent with it being governed by open system quantum dynamics. It is well understood that the D-Wave devices, just like any other quantum information processing device, must be error-corrected in order to overcome the effects of decoherence and control errors. While such error correction has already been demonstrated \cite{PAL:13}, our study focuses on the native performance of the device.

All annealing methods mentioned above are heuristic. They are not guaranteed to find the global optimum in a single annealing run, but only find it with a certain instance-dependent success probability $s\leq 1$. We determine the true ground state energy using an exact belief propagation algorithm \cite{dechter1999bucket}. We then perform at least $1000$ repetitions of the annealing for each instance, count how often the ground state has been found by comparing to the exact result, and use this to estimate the success probability $s$ for each problem instance.

The \emph{total annealing time} is defined as the time to perform $R$ annealing runs, where $R$ is the number of repetitions needed to find the ground state at least once with probability $p$: 
\begin{equation}
  R = \left\lceil
    \frac{\log (1-p)}{\log(1 - s)} 
    \right\rceil 
    \label{eq:repetitions}
\end{equation}
In order to reduce the effect of calibration errors on the DW2, it is advantageous to repeat  the annealing runs for several different encodings (``gauges'') of a problem instance. See Methods for details.

\section{Considerations when computing quantum speedup}

Let us first consider the subtleties of estimating the asymptotic scaling from small problem sizes $N$, and  inefficiencies at small problem sizes that can fake or mask a speedup. In the context of annealing methods the optimal choice of the annealing time turns out to be crucial for estimating asymptotic scaling.

\subsection{Asymptotic scaling: SA \textit{vs} SQA}
\label{sec:suboptimal}

%%%%%%%%%
%\input{fig1}
\begin{figure}[t]
\includegraphics[width=\columnwidth]{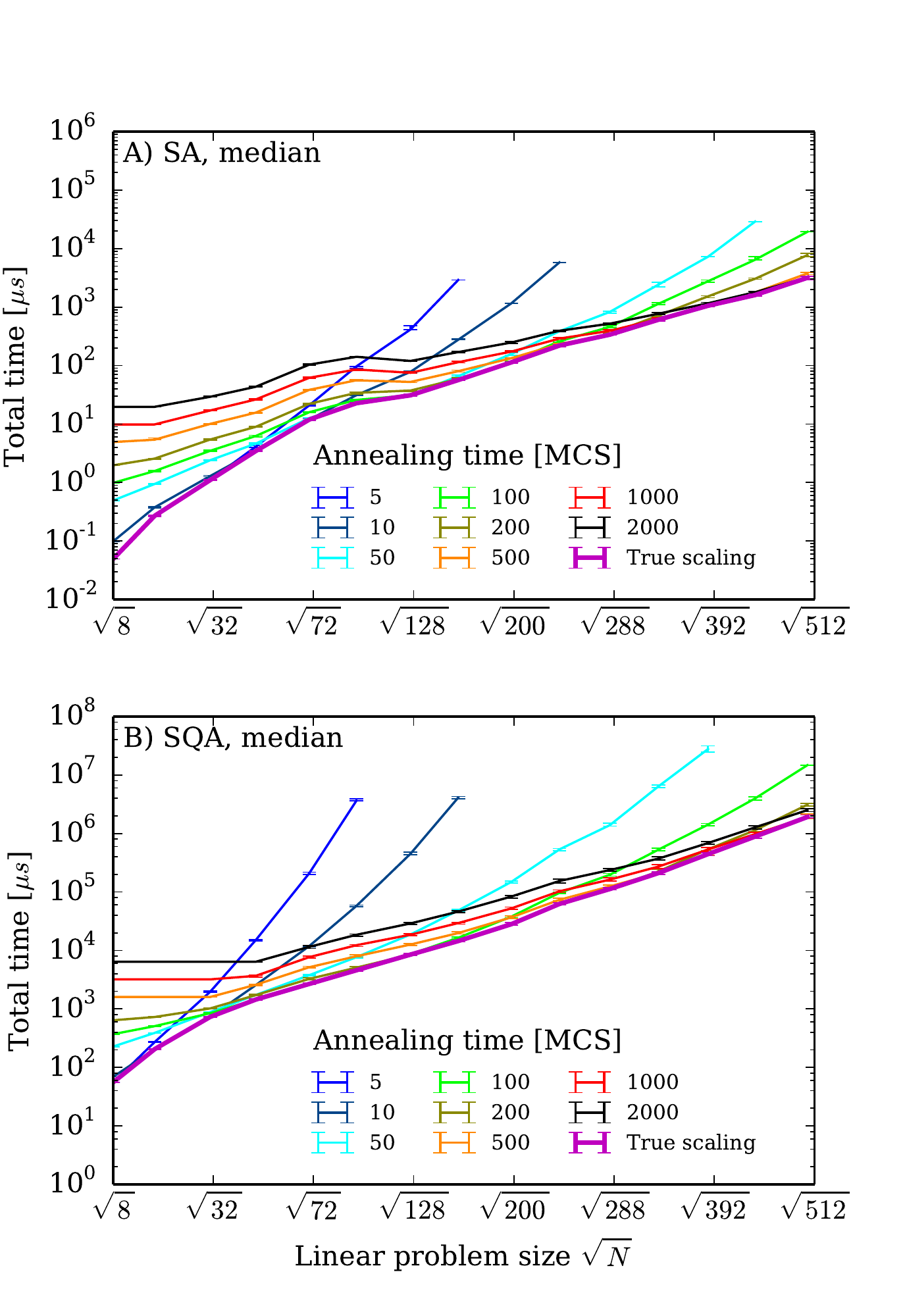}
\caption{{\bf Scaling of the typical time to find a solution at constant annealing time.} Shown is the typical (median)  time to find a ground state with 99\%  probability for spin glasses with $\pm1$ couplings and no local field. A) for SA, B) for SQA. The envelope of the curves at constant $t_a$, shown in red, corresponds to the minimal time at a given problem size $N$ and is relevant for discussion of the asymptotic scaling. Annealing times are given in units of Monte Carlo steps (MCS). One MCS corresponds to one update per spin. Note in particular that the slope for small $N$ is much flatter at large annealing time (e.g., $\textrm{MCS}=4000$) than that of the true scaling.}
\label{fig:medianfixed}
\end{figure}
\begin{figure}
\includegraphics[width=\columnwidth]{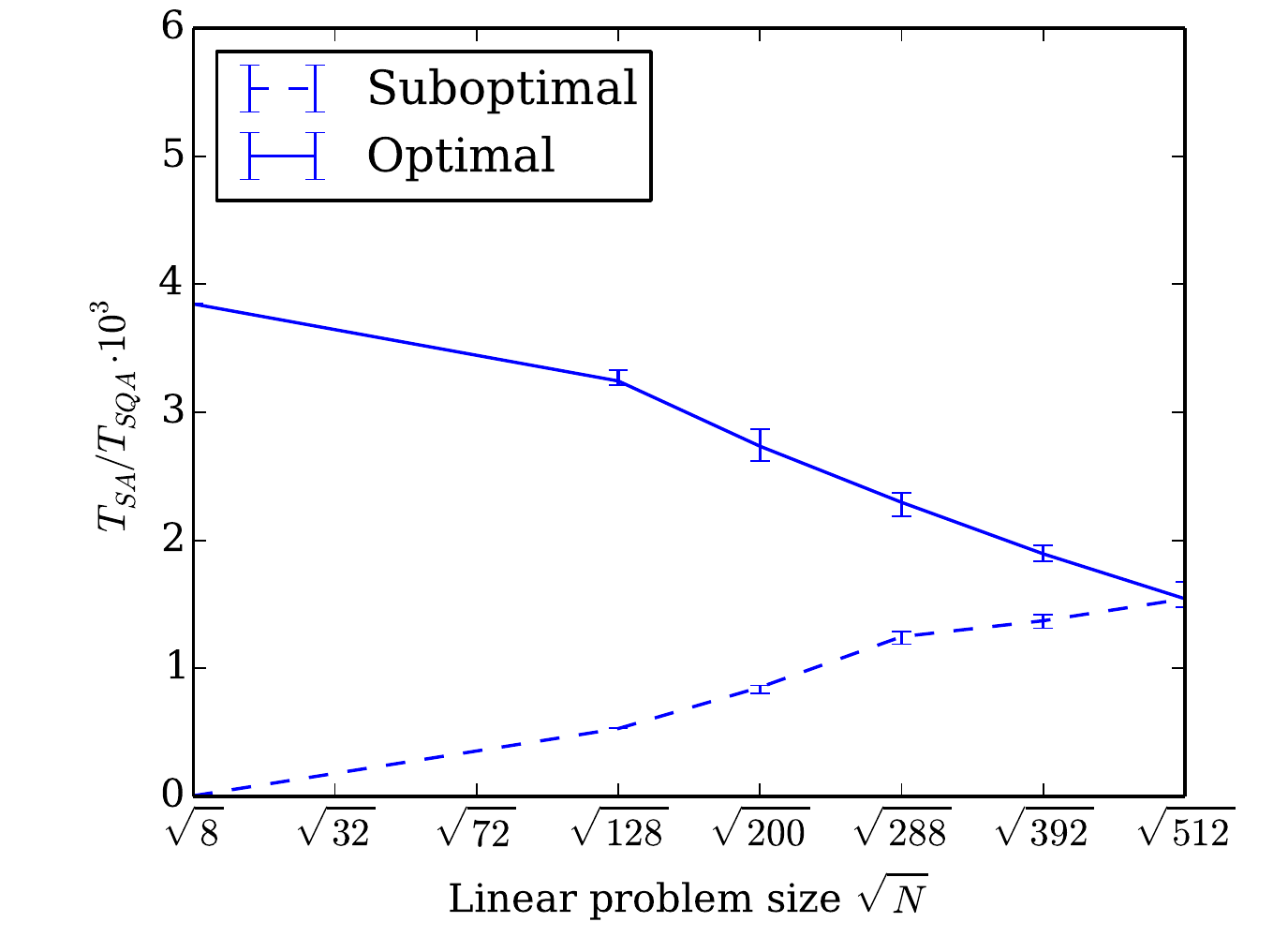}
\caption{{\bf Pitfalls when detecting speedup.} Shown is the speedup of SQA over SA, defined as the ratio of median time to find a solution with 99\% probability between SA and SQA. Two cases are presented: a) both SA and SQA run optimally (i.e., the ratio of the true scaling curves shown in Figure~\ref{fig:medianfixed}), and there is no asymptotic speedup (solid line). b) SQA is run suboptimally at small sizes by choosing a fixed large annealing time $t_a=10000$ MCS (dashed line). The apparent speedup is, however, due to suboptimal performance on small sizes and not indicative of the true asymptotic behavior given by the solid line, which displays a slowdown of SQA compared to SA.}
\label{fig:fakespeedup}
\end{figure}
%%%%%%%%%

To illustrate these issues we consider the time to solution using SA and SQA run at different \emph{fixed} annealing times $t_a$, independent of the problem size $N$. The problem instances we choose are random couplings of $J_{ij}=\pm1$ on each of the edges in a perfect Chimera graph of $L\times L$ unit cells, containing $N=8L^2$ spins (see Methods). We set the local fields $h_i=0$. Figure~\ref{fig:medianfixed} shows the scaling of the median total annealing time (over $1000$ different random instances) for both SA and SQA to find a solution with probability $p=0.99$. We observe that at constant $t_a$, as long as $t_a$ is long enough to find the ground state almost every time, the scaling of the total effort is at first relatively flat. The total effort then rises more rapidly, once one reaches problem sizes for which the chosen annealing time is too short and the success probabilities are thus low, requiring many repetitions. 

Figure~\ref{fig:medianfixed} demonstrates that no conclusion can be drawn from annealing (simulated or in a device) about the asymptotic scaling at fixed annealing times. It is misleading to conclude about the asymptotic scaling from the initial slow increase at constant $t_a$, and instead the \emph{optimal} annealing time $t_a^{\rm opt}$ needs to be found for each problem size $N$ \cite{ourpaper,opt-comment}. The lower envelope of the scaling curves (indicated in red in Figure~\ref{fig:medianfixed}) corresponds to the total effort at an optimal size-dependent annealing time $t_a^{\rm opt}(N)$ and can be used to infer the asymptotic scaling. In fact, the initial, relatively flat slope is due to suboptimal performance at small problem sizes $N$, and should therefore not be interpreted as speedup. To illustrate this we show in Figure~\ref{fig:fakespeedup} (solid line) the true ``speedup'' ratio of the scaling of SA and SQA (actually a slowdown), and a misleading, fake speedup (dashed line) due to using a constant and excessively long annealing time $t_a$ for SQA. Since the initial, slow increase of the total SQA effort at constant annealing time is a lower bound for the scaling of the true effort, the speedup \emph{slope} obtained from this data---which depends inversely on the SQA effort---is an \emph{upper bound}, as confirmed by Figure~\ref{fig:fakespeedup}. 

\subsection{Resource usage and speedup from parallelism}

A related issue is the scaling of hardware resources with problem size and parallelism in classical and quantum devices. To avoid mistaking a parallel speedup for a quantum speedup we need to scale hardware resources  (computational gates and memory) in the same way for the devices we compare, and employ these resources optimally. These considerations are not universal but need to be carefully applied for each comparison of a quantum algorithm and device to a classical one.

For a problem of size $N$, the DW2 uses only $N$ out of $512$ qubits and $\mc{O}(N)$ couplers and classical logical control gates to solve a spin glass instance with $N$ spin variables. We denote the time it needs to solve a problem by $T_{\textrm{DW}}(N)$. The classical simulated annealer (or simulated quantum annealer) running on a single classical CPU, on the other hand, uses fixed resources independent of problem size $N$, and we denote the time it requires to solve a problem by $T_{\rm SA}(N)$. We  consider here only the pure annealing times, as they are what is relevant for the asymptotic scaling rather than the readout or setup times, which scale subdominantly for large problems.

In order to avoid confusing quantum speedup with parallel speedup we thus consider as a classical counterpart to the DW2  a (hypothetical) special purpose parallel classical simulated annealing device, with the same hardware scaling as the DW2. Simulated annealing (and simulated quantum annealing) is perfectly parallelizable for the bipartite Chimera graphs realized by the DW2. The reason is that one Monte Carlo step (consisting of one attempted update per spin) can be performed in constant time, since all spins in each of the two sublattices can be updated simultaneously. The time to solve a problem on this equivalent classical device, denoted by $T_{\rm C}(N)$, is thus related to the time $T_{\rm SA}(N)$ taken by a simulated annealer using a fixed-size classical CPU by
\begin{equation}
T_{\rm C}(N) \propto \frac{1}{N}T_{\rm SA}(N) ,
\end{equation}
since the latter needs time $O(N)$ for one Monte Carlo step, while the former performs it in constant time.

The {\em quantum} part of speedup is then estimated by comparing the times required by two devices with the same hardware scaling, giving
\begin{equation}
S(N) = \frac{T_{\textrm{C}}(N)}{T_{\textrm{DW}}(N)} \propto  \frac{T_{\textrm{SA}}(N)}{T_{\textrm{DW}}(N)} \frac{1}{N}.
\label{eq:parallelspeedup1}
\end{equation}
The factor $1/N$ in the speedup calculation thus discounts for the intrinsic \emph{parallel speedup} of the analog device whose hardware resources scale as $N$. See Methods for an alternative derivation that leads to the same results (up to subleading corrections) by using a fixed size device efficiently.

\section{Performance of  D-Wave Two versus SA and SQA}
\label{sec:DW2-performance}

\subsection{Comparing  devices}

%How to define and compare speedup between two devices---such as the DW2 and simulated classical and quantum annealers---depends on the goal of the study. 

If the goal is to compare the performance of devices as optimizers, then one is interested in solving \emph{almost all} problem instances. In this case we should run the devices in such a way that all but a small fraction of the problems can be solved.
This will lead to a speedup defined as the ratio of the  quantiles (``R of Q") of the time to solution, with an emphasis on the high quantiles, which we discuss in Sec.~\ref{sec:roq}. 
A complementary question is to ask whether a device exhibits better performance than another for \emph{some} problems. To answer this question we compare the time to solution individually for each problem instance. We  then consider the quantiles of the ratio (``Q of R") of the time to solution, and discuss this approach in Sec.~\ref{sec:qor}.

A complementary distinction is that between \emph{wall-clock time}, denoting the full time to solution, and the \emph{pure annealing time}. Wall-clock time is the total time to find a solution and is the relevant quantity when one is interested in the performance of a device for applications and has been used in Ref. \cite{McGeoch}. It includes the setup, cooling, annealing and readout times on the DW2, and the setup, annealing and measurement time for the classical annealing codes. The pure annealing time is simply $Rt_a$, where $R$ is the number of repetitions and $t_a$ the time used for a single annealing run. It is the relevant quantity when one is interested in the intrinsic physics of the annealing processes and in scaling to larger problem sizes on future devices. We discuss both wall-clock and pure annealing times below.

\subsection{Problem instances}
\label{sec:problems}
The family of problem instances we use for our benchmarking tests employ couplings $J_{ij}$ on all edges of $N=8LL'$-vertex subgraphs of the Chimera graph of the DW2, comprising $L\times L'$ unit cells, with $L,L'\in\{1,\dots,8\}$. We set the fields $h_i=0$ since nonzero values of the fields $h_i$ destroy the spin glass phase that exists at zero field, thus making the instances easier \cite{AT}. We choose the values of the couplings $J_{ij}$ from $2r$ discrete values  $\{n/r\}$, with $n \in \{-r, -r-1, \dots, -1, 1, \dots, r-1, r\}$, and call $r$ the ``range". Thus when the range $r=1$ we only pick values $J_{ij}=\pm 1$. This choice is the least susceptible to calibration errors of the device, but the large degeneracy of the ground states in these cases makes finding a ground state somewhat easier. At the opposite end we consider $r=7$, which is the upper limit given the four bits of accuracy of the couplings in the DW2. These problem instancess are harder since there are fewer degenerate minima, but they also suffer more from calibration errors in the device. In the Supplementary Material we present additional results for $r=3$.

\subsection{Performance as an optimizer: comparing the scaling of hard problem instances}
\label{sec:roq}

\subsubsection{Pure annealing time}
\label{sec:pure-t_a}

%%%%%%%%%
%\input{fig3}
\begin{figure*}[ht]
\includegraphics[width=1.8\columnwidth]{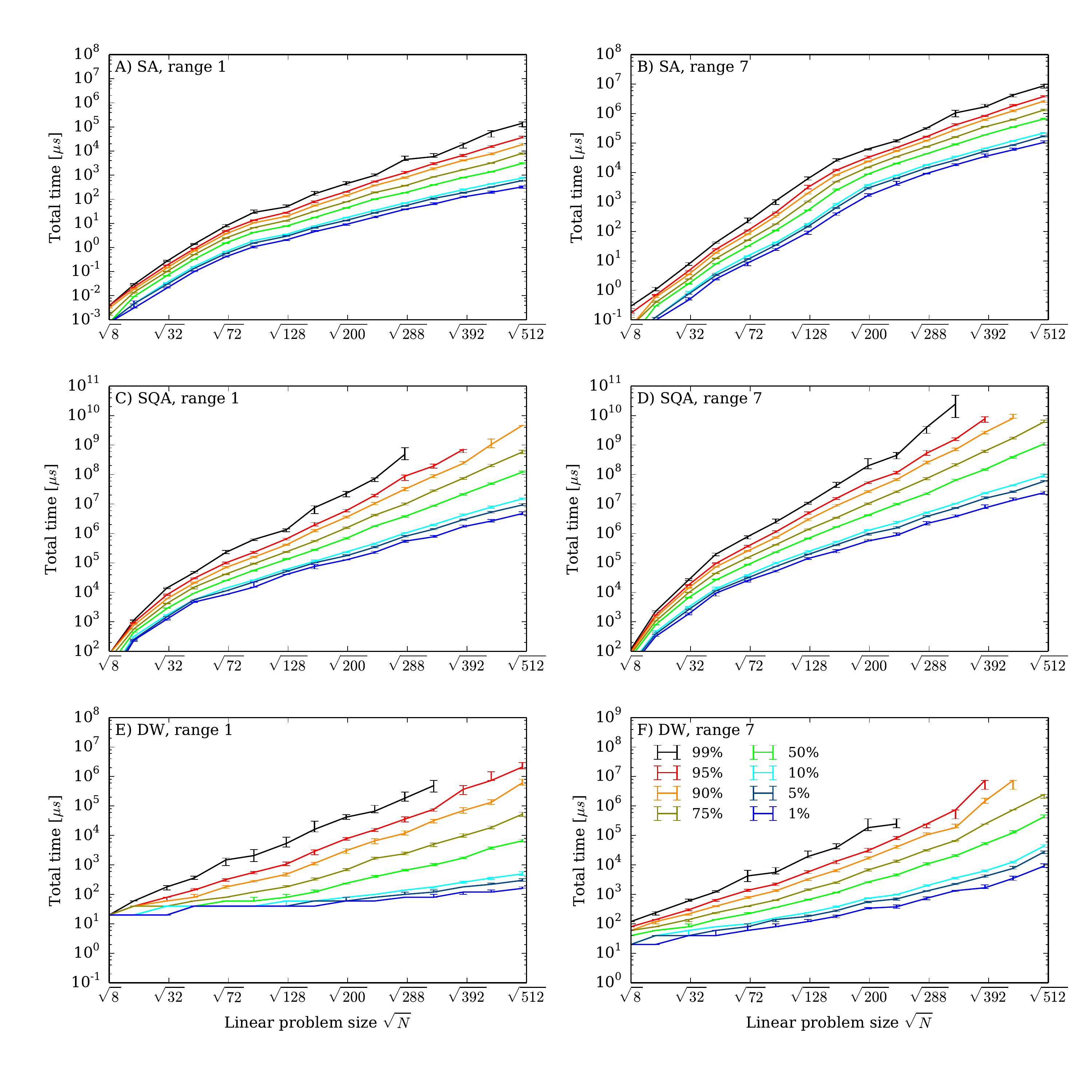} \label{fig:scalingraw7}
\caption{{\bf Scaling of time to solution for the ranges $r=1$ (panels A, C and E) and $r=7$ (panels B, D and F).} Shown is the scaling of the time to find the ground state at least once with a probability $p=0.99$ for various quantiles of hardness, for A,B) simulated annealing (SA), C,D) simulated quantum annealing (SQA) and E,F) the DW2. The SA and SQA data is obtained by running the simulations at an optimized annealing time for each problem size. The DW2 annealing time of $20\mu s$ is the shortest possible. 
Note the different vertical axis scales, and that both the DW2 and SQA have trouble solving the hardest instances for the large problem sizes, as indicated by the terminating lines for the highest quantiles. More than the maximum number of of repetitions (10000 for SQA, at least 32000 for DW2) of the annealing we performed would be needed to find the ground state in those cases.
}
\label{fig:scalingraw}
\end{figure*}
%%%%%%%%%

We start our analysis by focusing on pure annealing times and show in Figure~\ref{fig:scalingraw} the scaling of the time to find the ground state at least once with probability $p=0.99$ for various quantiles, from the easiest instances (1\%) to the hardest (99\%), for two different ranges. Since we do not \textit{a priori} know the hardness of a given problem instance we have to assume the worst case and perform a sufficient number of repetitions $R$ to be able to solve even the hardest problem instances. Hence the scaling for the selected high quantile will apply to \emph{all} problem instances we run on the optimizer.

In all three cases (SA, SQA, DW2) we observe, for sufficiently large $N$, that the total time to solution scales with $\exp(c\sqrt{N})$, as reported previously for SA and SQA \cite{ourpaper}. The origin of the $\sqrt{N}$ exponent is well understood for exact solvers as reflecting the treewidth of the Chimera graph (see Methods and Ref.~\cite{Choi2}), and a similar scaling is observed here for the heuristic algorithms.
While the SA and SQA codes were run at an optimized annealing time for each problem size $N$, the DW2 has a minimal annealing time of $t_a=20\mu s$, which is longer than the optimal time for all problem sizes (see Methods).
Therefore the observed slope of the DW2 data 
should only be taken as a lower bound for the asymptotic scaling. Even so, we observe similar scaling for the classical codes and  on DW2.
  
\subsubsection{The ratio of quantiles}
 
%%%%%%%%%
%\input{fig4}
\begin{figure}
\includegraphics[width=\columnwidth]{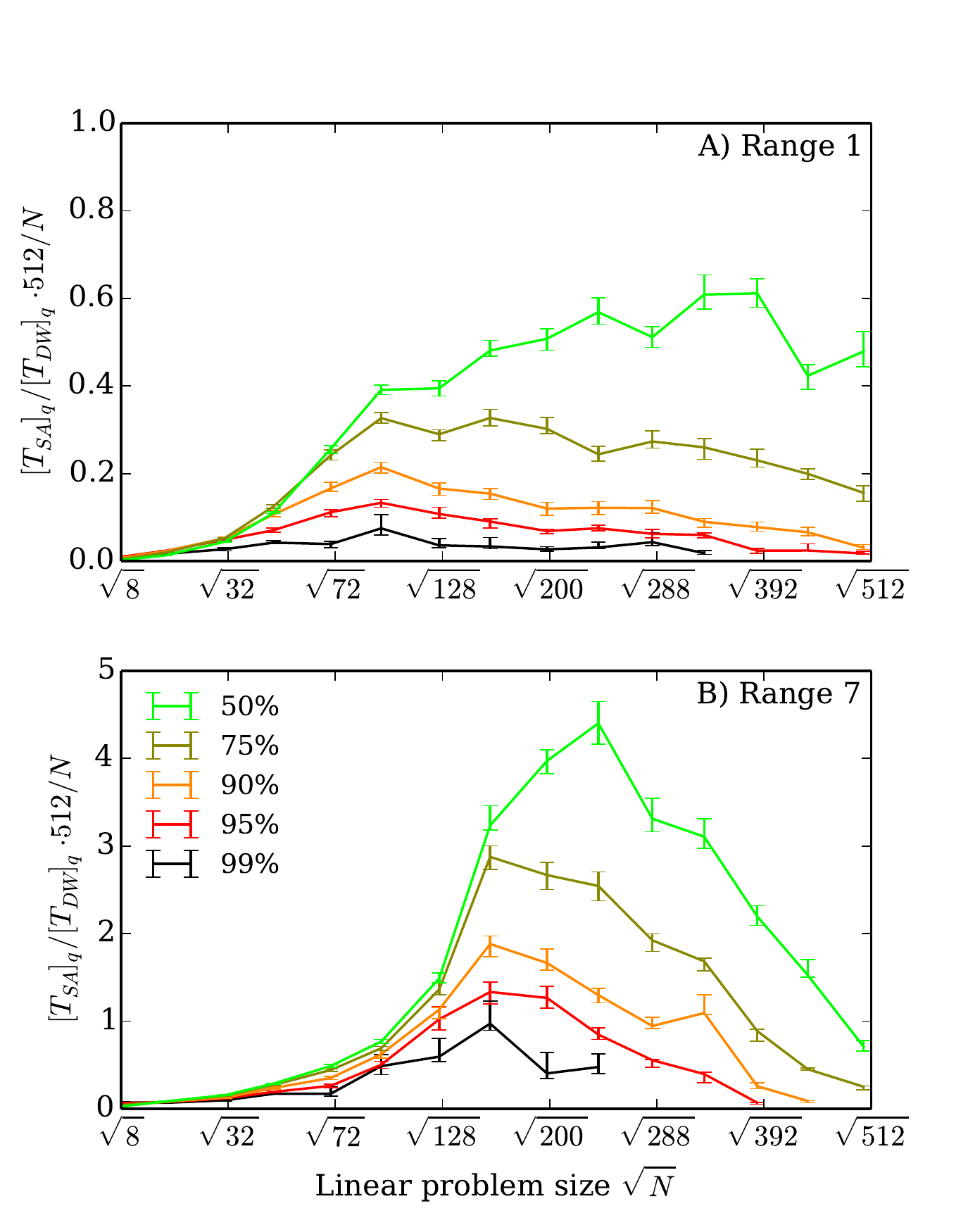}
\caption{{\bf Speedup for ratio of quantiles for the DW2 compared to SA.} A) For instances with range $r=1$. B) For instances with range $r=7$. Shown are curves from the median ($50$th quantile) to the $99$th quantile. $16$ gauges were used. In these plots we multiplied Eq.~\eqref{eq:S_q} by $512$ so that the speedup value at $N=512$ directly compares one DW2 processor against one classical CPU.}
\label{fig:qorspeedup1}
\end{figure}
%%%%%%%%%

With algorithms such as SA or quantum annealing, where the time to solution depends on the problem instance, it is often not possible (and usually irrelevant) to experimentally find the hardest problem instance. It is preferable to decide instead for which fraction of problem instances one wishes to find the ground state, which then defines the relevant quantile. If we target $q$\% of the instances then we should consider the $q$th percentile in the scaling plots shown in Figure~\ref{fig:scalingraw}. The appropriate speedup quantity is then the \emph{ratio of these quantiles}. Denoting a quantile $q$ of a random variable $X$ by $[X]_q$ we can define this as
\begin{equation}
S^{\textrm{RofQ}}_q(N) = \frac{[T_{\textrm{C}}(N)]_q}{[T_{\textrm{DW}}(N)]_q} \propto \frac{[T_{\textrm{SA}}(N)]_q}{[T_{\textrm{DW}}(N)]_q}  \frac{1}{N}.
\label{eq:S_q}
\end{equation}
Plotting this quantity for the DW2 \textit{vs} SA in Figure~\ref{fig:qorspeedup1}
we find no evidence for a limited quantum speedup in the interesting regime of large $N$. That is, while for all quantiles, and for both ranges  (with the exception of the $50$th quantile and $r=1$), the initial slope is positive, when $N$ becomes large enough we observe a turnaround and eventually a negative slope, showing that SA outperforms the DW2.

Taking into account that (as discussed in Sec.~\ref{sec:suboptimal}) due to the fixed suboptimal annealing times the speedup defined in Eq.~\eqref{eq:S_q} is an \emph{upper bound}, we conclude that the DW2 does not exhibit a speedup over SA for this particular benchmark. 

\subsubsection{Wall-clock time}

%%%%%%%%%
%\input{fig5}
\begin{figure}
\includegraphics[width=\columnwidth]{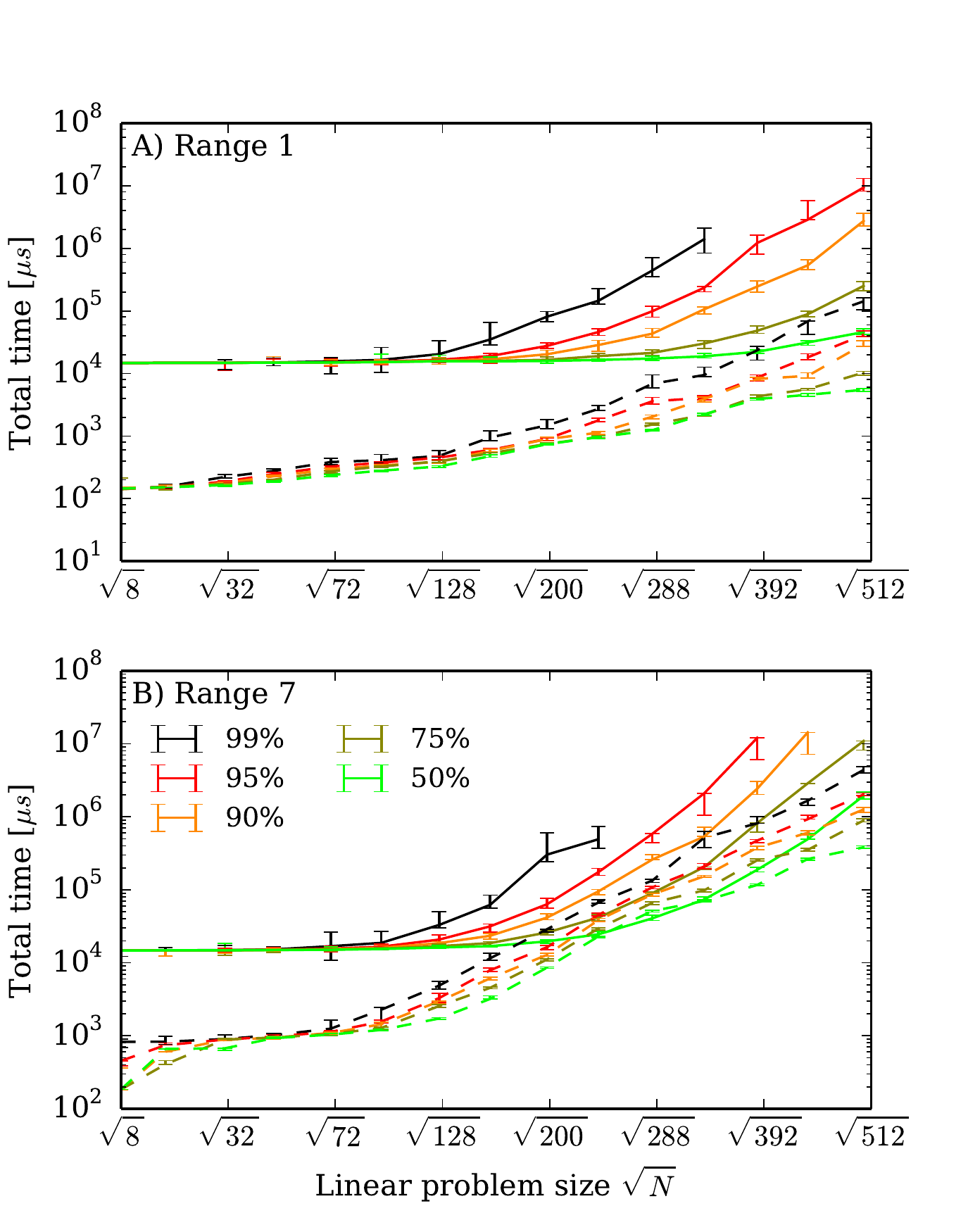}
\caption{{\bf Comparing wall-clock times} A comparison of the wall-clock time to find the solution with probability $p=0.99$ for SA running on a single CPU (dashed lines) compared to the DW2 (solid lines) using $16$ gauges. A) for range $r=1$, B) for range $r=7$. Shown are curves from the median ($50$th quantile) to the $99$th quantile. The large constant programming overhead of the DW2 masks the exponential increase of time to solution that is obvious in the plots of pure annealing time. Results for a single gauge are shown in the Supplementary Material.}
\label{fig:wall-clock1}
\end{figure}
%%%%%%%%%

While not as interesting from a complexity theory point of view, it is instructive to also compare wall-clock times for the above benchmarks, as we do in Figure~\ref{fig:wall-clock1}. We observe that the DW2 performs similarly to SA run on a single classical CPU, for sufficiently large problem sizes and at high range values. Note that the large constant programming overhead of the DW2 masks the exponential increase of time to solution that is obvious in the plots of pure annealing time.

%%%%%%%%%
%\input{fig6}
\begin{figure*}[ht]
\centering
\includegraphics[width=0.95\textwidth]{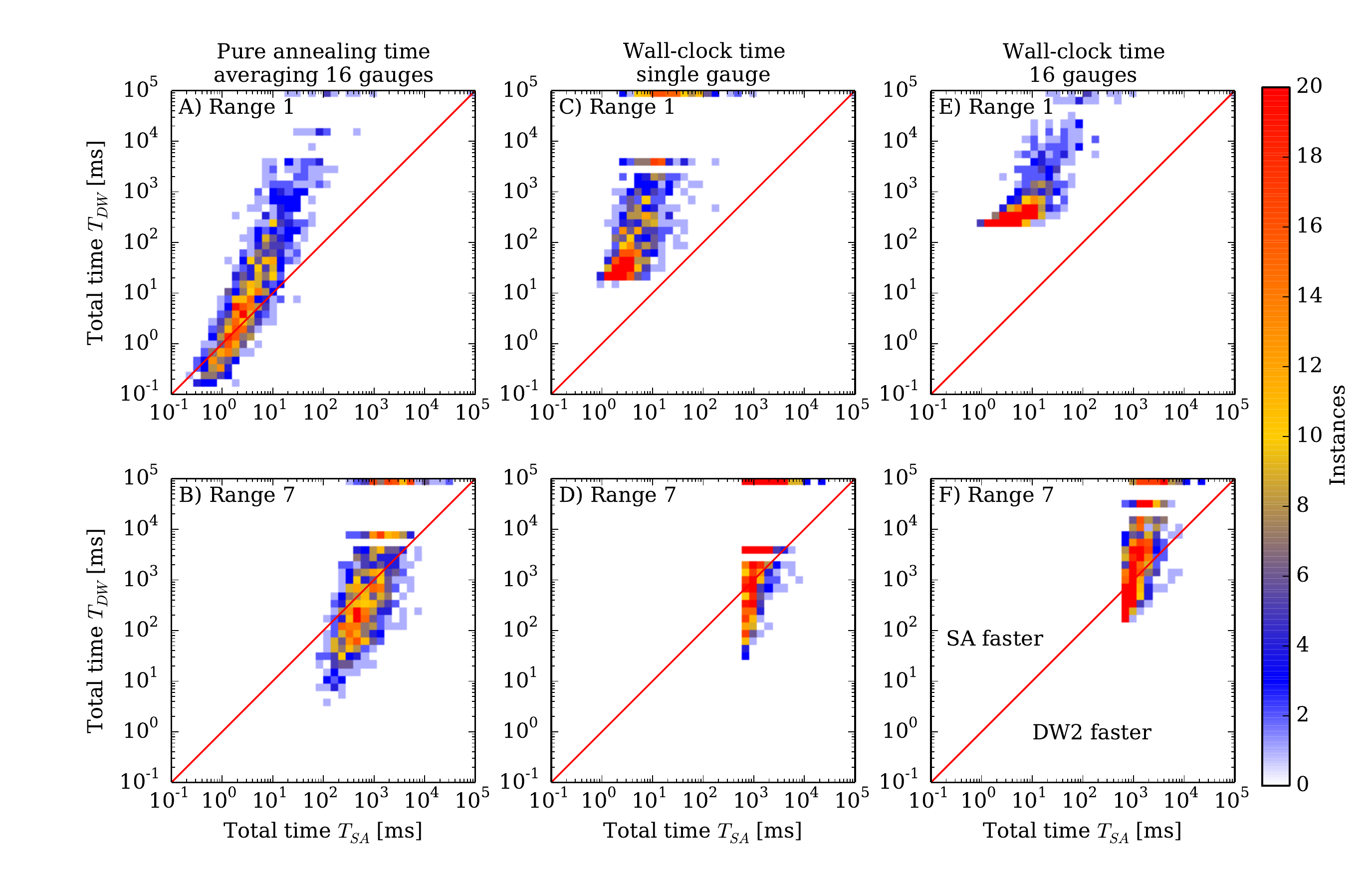} \label{fig:ratiosannealing}
\caption{{\bf Instance-by-instance comparison of annealing times and wall-clock times.} Shown is a scatter plot of the pure annealing time for the DW2 compared to a simulated classical annealer (SA) using an average over $16$ gauges on the DW2. A)  DW2 compared to SA for $r=1$, B) DW2 compared to SA for $r=7$.  The color scale indicates the number of instances in each square. Instances below the diagonal red line are faster on the DW2, those above are faster classically. Instances for which the DW2 did not find the solution with 10000 repetitions per gauge are shown at the top of the frame (no such instances were found for SA). 
Panels C) and D)  show  wall-clock times using a single gauge on the DW2.
Panels E) and F) show the wall-clock time for DW2 using $16$ gauges. $N=503$ in all cases.}
\label{fig:ratiosannealing}
\end{figure*}
%%%%%%%%%

\subsection{Instance-by-instance comparison}
\label{sec:qor}

\subsubsection{Total time to solution}

We now focus on the question of whether the DW2 exhibits a limited quantum speedup for some fraction of the instances of our benchmark set. To this end we perform individual comparisons for each instance and show in Figure~\ref{fig:ratiosannealing}A-B the ratios of time to solution between the DW2 and SA, considering only the pure annealing time. We find a wide scatter, which is not surprising since we previously found that DW1 performs like a simulated quantum annealer, but correlates less well with a simulated classical annealer \cite{ourpaper}. We find that while the DW2 is sometimes up to $10\times$ faster in pure annealing time, there are many cases where it is $\geq 100\times$ slower. 

Considering the wall-clock times, the advantage of the DW2 seen in Figure~\ref{fig:ratiosannealing}A-B for some instances tends to disappear, since it is penalized by the need for programming the device with multiple different gauge choices (see Methods). Figure~\ref{fig:ratiosannealing}C-D shows that for one gauge choice there are some instances, for $r=7$, where the DW2 is faster, but many instances where it never finds a solution. Using $16$ gauges the DW2 finds the solution in most cases, but is always slower than the classical annealer on a classical CPU for $r=1$, as can be seen in Figure~\ref{fig:ratiosannealing}E-F. For $r=7$ the DW2 is sometimes faster than a single classical CPU. Overall, the performance of the DW2 is better for $r=7$ than for $r=1$, and comparable to SA only when just the pure annealing time is considered. The difference to the results of Ref. \cite{McGeoch} is due to the use of optimized classical codes using a full CPU in our comparison, as opposed to the use of generic optimization codes using only a single CPU core in Ref. \cite{McGeoch}.

%%%%%%%%%
%\input{fig7}
\begin{figure}
\includegraphics[width=\columnwidth]{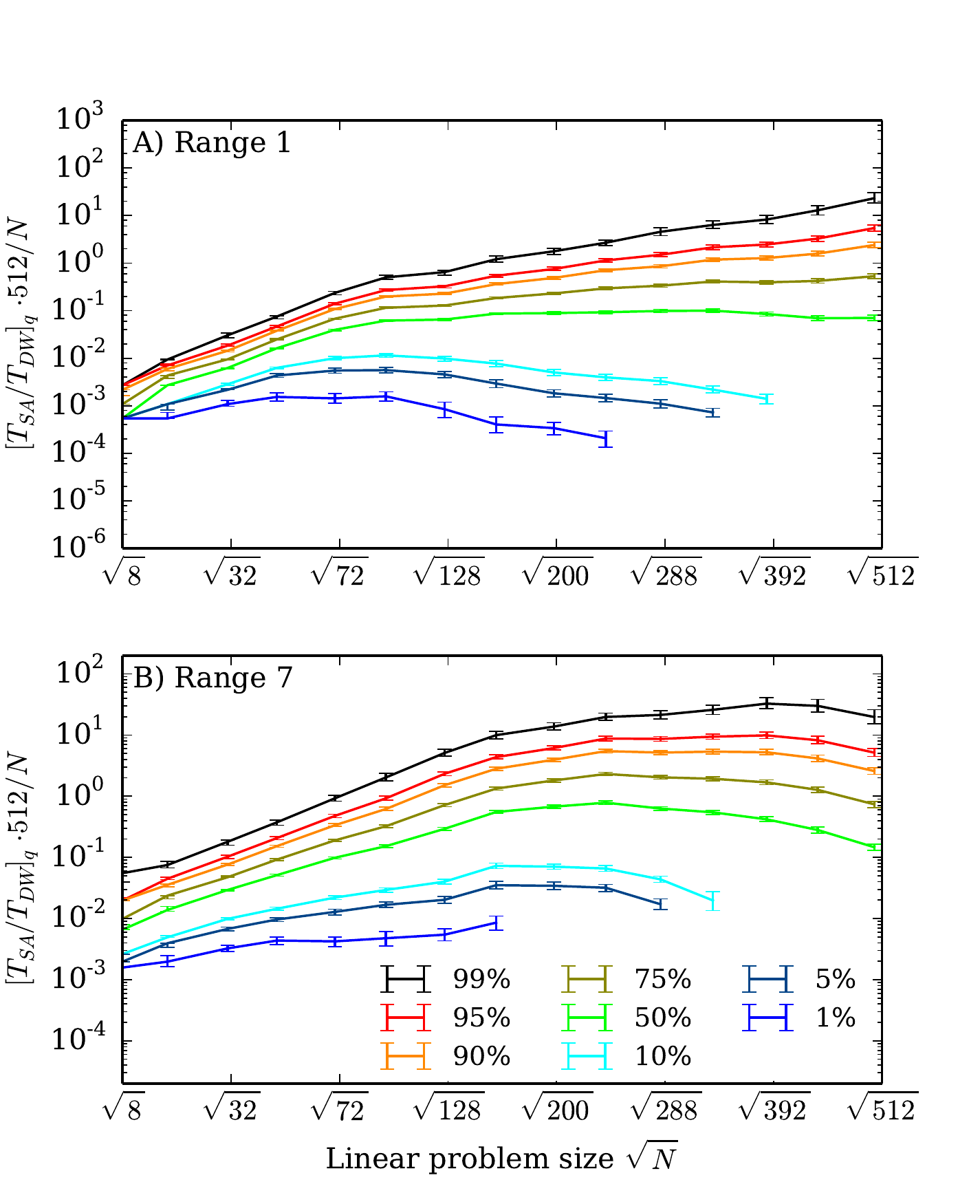}
\caption{{\bf Speedup for quantiles of the ratio} of the DW2 compared to SA, for A) $r=1$,  B) $r=7$. No asymptotic speedup is visible for any of the quantiles at $r=7$, while some evidence of a limited quantum speedup (relative to SA) is seen for quantiles higher than the median at $r=1$. As in Figure~\ref{fig:qorspeedup1} we multiplied Eq.~\eqref{eq:SQoR} by $512$ so that the speedup value at $N=512$ directly compares one DW2 processor against one classical CPU.}
\label{fig:qorspeedup7}
\end{figure}
%%%%%%%%%

\subsubsection{Quantiles of ratio}

Comparisons of the absolute time to solution are of limited importance compared to the real question of scaling, which can give insight into the behavior of future devices that can solve larger problems. In Section \ref{sec:roq} we did not find evidence for a limited quantum speedup when considering all instances. Now we consider instead whether there is such a speedup for a subset of problem instances. 
To this end we study the scaling of the ratios of the time to solution for individual instances, and display in Figure~\ref{fig:qorspeedup7} the scaling of various \emph{quantiles of the ratio} 
\begin{equation}
S_q^{\textrm{QofR}}(N) = \left[\frac{T_{\textrm{C}}(N)}{T_{\textrm{DW}}(N)}\right]_q \propto \left[\frac{T_{\textrm{SA}}(N)}{T_{\textrm{DW}}(N)}\frac{1}{N}\right]_q  .
\label{eq:SQoR}
\end{equation}
For $r=7$ all the quantiles bend down for sufficiently large $N$, so that there is no evidence of a limited quantum speedup. Yet, now there seems to be an indication of such a speedup compared to SA in the high quantiles for $r=1$. However, for the reasons discussed in Sec.~\ref{sec:suboptimal}, one must be careful not to overinterpret this as solid evidence for a speedup since the instances contributing here are not run at the optimal annealing time. Moreover, as discussed in the Supplementary Material, we find no evidence of a limited quantum speedup for $r=3$.
%, so that it seems unlikely that calibration problems alone at the high range of coupler accuracy ($r=7$) are responsible for the lack of a speedup.
Thus, while perhaps encouraging from the perspective of a search for a (limited) quantum speedup, more work is needed to establish that the $r=1$ result persists for those instances for which one can be sure that the annealing time is optimal. 

\subsection{Arguments for and against a speedup on the DW2}

Let us consider in some more detail the speedup results discussed above. We have argued that the apparent limited quantum speedup seen in the $r=1$ results of Figure~\ref{fig:qorspeedup7} must be treated with care due to the suboptimal annealing time. 
It might then be tempting to argue that, strictly speaking, the comparison with suboptimal-time instances cannot be used for claiming a slowdown either, i.e., that we simply cannot infer how the DW2 will behave for optimal-time instances by basing the analysis on suboptimal times only. 

However, let us make the assumption that, along with the total time, the optimal annealing time $t_a^{\textrm{opt}}(N)$ also grows with problem size $N$. This assumption is supported by the SA and SQA data shown in Figure~\ref{fig:medianfixed}, and is plausible as long as the growing annealing time does not become counterproductive due to coupling to the thermal bath \cite{PhysRevLett.95.250503}. By definition,  $T_{\textrm{DW}}(N,t_a^{\textrm{opt}}(N)) \leq T_{\textrm{DW}}(N,t_a)$, where we have added the explicit dependence on the annealing time, and $t_a$ is a fixed annealing time. Thus
\begin{align}
S(N) &= \frac{T_{\textrm{C}}(N)}{T_{\textrm{DW}}(N,t_a)}\frac{1}{N}  \\
&\leq \frac{T_{\textrm{C}}(N)}{T_{\textrm{DW}}(N,t_a^{\textrm{opt}}(N))}\frac{1}{N} = S^{\textrm{opt}}(N) \notag.
\end{align}
Under our assumption, $t_a^{\textrm{opt}}(N) < t_a $ for small $N$, but for sufficiently large $N$ the optimal annealing time grows so that $t_a^{\textrm{opt}}(N) \geq t_a$. Thus there must be a problem size $N^*$ at which $t_a^{\textrm{opt}}(N^*) = t_a$, and hence at this special problem size we also have $S(N^*) = S^{\textrm{opt}}(N^*)$. However, as mentioned in Section~\ref{sec:pure-t_a}, the minimal annealing time of $20\mu s$ is longer than the optimal time for all problem sizes (see Supplementary Material), i.e., $N^*>503$ in our case. Therefore, if $S(N)$ is a decreasing function of $N$ for sufficiently large $N$, as we indeed observe in all our ``R of Q" results (recall Figure \ref{fig:qorspeedup1}), then since $S^{\textrm{opt}}(N) \geq S(N)$ \emph{and} $S(N^*) = S^{\textrm{opt}}(N^*)$, it follows that $S^{\textrm{opt}}(N)$ too must be a decreasing function for a range of $N$ values, at least until $N^*$. This shows that the slowdown conclusion holds also for the case of optimal annealing times. 

For the instance-by-instance comparison (``Q of R"), no such conclusion can be drawn for the subset of instances (at $r=1$) corresponding to the high quantiles where $S_q^{\textrm{QofR}}(N)$ is an \emph{increasing} function of $N$. This limited quantum speedup may or may not persist for larger problem sizes or if optimal annealing times are used. 

\section{Discussion}

In this work we have discussed challenges in properly defining and assessing quantum speedup, and used comparisons between a DW2 and simulated classical and quantum annealing to illustrate these challenges. \emph{Strong} or \emph{provable} quantum speedup, implying speedup of a quantum algorithm or device over \emph{any} classical algorithm, is an elusive goal in most cases and one thus usually defines \emph{quantum speedup} as a speedup compared to the best available classical algorithm. We have introduced the notion of \emph{limited quantum speedup}, referring to a more restricted comparison to ``corresponding'' classical algorithms solving the same task, such as a quantum annealer compared to a classical annealing algorithm.

Quantum speedup is most easily defined and detected in the case of an exponential speedup, where the details of the quantum or classical hardware do not matter since they only contribute subdominant polynomial factors. In the case of an unknown or a polynomial quantum speedup one must be careful to fairly compare the classical and quantum devices, and, in particular, to scale hardware resources in the same manner. Otherwise \emph{parallel speedup} might be mistaken for (or hide) quantum speedup.

An experimental determination of quantum speedup suffers from the problem that all measurements are limited to finite problem sizes $N$, while we are most interested in the asymptotic behavior for large $N$. To arrive at a reliable extrapolation it is advantageous to focus the scaling analysis on the part of the execution time that becomes dominant for large problem sizes $N$, which in our example is the pure annealing time, and not the total wall-clock time. For each problem size we furthermore need to ensure that neither the quantum device nor the classical algorithm are run suboptimally, since this might hide or fake quantum speedup.

If the time to solution depends not only on the problem size $N$ but also on the specific problem instance, then one needs to carefully choose the relevant quantity to benchmark. We argued that in order to judge the performance over many possible inputs of a randomized benchmark test, one needs to study the high quantiles, and define speedup by considering the ratio of the quantiles of time to solution. If, on the other hand, one is interested in finding out whether there is a speedup for some subset of problem instances, then one can instead perform an instance-by-instance comparison by focusing on the quantiles of the ratio of time to solution.

We note that it is not yet known whether a quantum annealer or even a perfectly coherent adiabatic quantum optimizer can exhibit (limited) quantum speedup at all  \cite{exception-comment}, although there are promising indications from simulation \cite{Santoro} and experiments on spin glass materials \cite{Brooke1999}. Experimental tests will thus be important. We chose to focus here on the benchmark problem of random zero-field Ising problems parametrized by the range of couplings. 
We did not find evidence of limited quantum speedup for the DW2 relative to simulated annealing in our particular benchmark set when we considered the ratio of quantiles of time to solution, which is the relevant quantity for the performance of a device as an optimizer. We note that random spin glass problems, while an interesting and important physics problem, may not be the most relevant benchmark for practical applications, for which other benchmarks may have to be studied.

When we focus on subsets of problem instances in an instance-by-instance comparison, we observe a possibility for a limited quantum speedup for a fraction of the instances  \cite{hybrid-comment}. However, since the DW2 runs at a suboptimal annealing time for most of the corresponding problem instances, the observed speedup may be an artifact of attempting to solve the smaller problem sizes using  an excessively long annealing time. This difficulty can only be overcome by fixing the issue of suboptimal annealing times, e.g., by finding problem classes for which the annealing time is demonstrably already optimal. 

There are several candidate explanations for the absence of a clear quantum speedup in our tests. Perhaps quantum annealing simply does not provide any advantages over simulated (quantum) annealing or other classical algorithms for the problem class we have studied \cite{katzgraber}; or, perhaps, the noisy implementation in the DW2 cannot realize quantum speedup and is thus not better than classical devices. Alternatively, a speedup might be masked by calibration errors, improvements might arise from error correction \cite{PAL:13}, or other problem classes might exhibit a speedup \cite{MAX2SAT-comment}. Future studies will probe these alternatives and aim to determine whether one can find a class of problem instances for which an unambiguous speedup over classical hardware can be observed.

%\clearpage
\bigskip
\bigskip

%%%%%%%%%

\noindent {\textbf {METHODS}\\\\
%%%%%%%%%%%%%%%%%%%%%%%%%
{\small
\noindent \textbf{Simulated annealing.}
%\label{sec:simulated-annealing}
Simulated annealing \cite{Kirkpatrick1983} performs a Monte Carlo simulation on the model of Eq.~\eqref{eq:H}, starting from a random initial state at high temperature. During the course of the simulation the temperature is lowered towards zero. At the end of the annealing schedule, at low temperature, the spin configuration of the system ends up in  in a local minimum. By repeating the simulation many times one may hope to find the global minimum. More specifically, SA is performed by sequentially iterating  through all spins and proposing to flip them based on a Metropolis algorithm using the Boltzmann weight of the configuration at finite temperature. During the annealing schedule we linearly increase the inverse temperature over time from an initial value of $\beta=0.1$ to a final value of $\beta=3r$.

For the case of $\pm1$ couplings ($r=1$), and  for $r=3$ we use a highly optimized multispin-coded algorithm based on Refs.~\cite{J.Stat.Phys.44.985,Comput.Phys.Commun.59.387}. This algorithm performs updates on $64$ copies in parallel, updating all at once. For the $r=7$ simulations we use a code optimized for bipartite lattices \cite{sapaper}. Implementations of the simulated annealing codes are available in Ref.~\cite{sapaper}. We used the code {\tt an\_ms\_r1\_nf} for $r=1$, the code {\tt an\_ms\_r3\_nf} for $r=3$ and the code {\tt an\_ss\_ge\_nf\_bp } for $r=7$.\\

\noindent \textbf{Quantum annealing.}
To perform quantum annealing one maps the Ising variables $\sigma_i^z$ to Pauli $z$-matrices and adds a transverse magnetic field in the $x$-direction to induce quantum fluctuations, thus obtaining the time-dependent quantum Hamiltonian 
\begin{equation}
\label{eq:Hquantum}
H(t) = -A(t) \sum_i \sigma_i^x +B(t) H_{\rm Ising}\ , \quad t\in[0,t_a]\ .
\end{equation}
The annealing schedule starts at time $t=0$ with just the transverse field term (i.e., $B(0)=0$) and $A(0)\gg k_B T$, where $T$ is the temperature, which is kept constant. The system is then in a simple quantum state with (to an excellent approximation) all spins aligned in the $x$ direction, corresponding to a uniform superposition over all $2^N$ computational basis states (products of eigenstates of the $\sigma_i^z$). During the annealing process the problem Hamiltonian magnitude $B(t)$ is increased and the transverse field $A(t)$ is decreased, ending with $A(t_a)=0$, and couplings much larger than the temperature: $B(t_a) \max(\max_{ij}|J_{ij}|,\max_i|h_i|) \gg k_BT$. 
At this point the system will again be trapped in a local minimum, and by repeating the process one may hope to find the global minimum. Quantum annealing can be viewed as a finite-temperature variant of the adiabatic quantum algorithm \cite{farhi}.\\

\noindent \textbf{Simulated quantum annealing.}
%\label{sec:discr-time-quant}
Simulated quantum annealing (SQA) \cite{sqa1,Santoro} is an annealing algorithm based on discrete-time path-integral quantum Monte Carlo simulations of the transverse field Ising model, following the above annealing schedule at a constant low temperature, but using Monte Carlo dynamics instead of the open system evolution of a quantum system. This amounts to sampling the world line configurations of the quantum Hamiltonian \eqref{eq:Hquantum} while slowly changing the couplings.  The algorithm we used is similar to that of Ref.~\cite{PhysRevB.66.094203}, but uses cluster updates along the imaginary time direction, typically with $64$ time slices. Our annealing schedule is linear, as shown in Figure~\ref{fig:schedule}B): the Ising couplings are ramped up linearly while the transverse field is ramped down linearly over time.\\

\noindent \textbf{The layout of the D-Wave Two Vesuvius chip.}
The Chimera graph of the DW2 used in our tests is shown in Figure~\ref{fig:chimera}. Each unit cell is a balanced $K_{4,4}$ bipartite graph. In the ideal Chimera graph (of $512$ qubits) the degree of each vertex is $6$. 
For the scaling analysis we considered $L\times  L'$ rectangular sub-lattices of the Chimera graph, and restricted our simulations and tests on the DW2 to the subset of functional qubits within these subgraphs.
More generally the $N=2cL^2$-vertex Chimera graph comprises an $L\times L$ grid of $K_{c,c}$ unit cells, and the (so-called {\sc TRIAD}) construction of Ref.~\cite{Choi2} can be used to embed the complete $L$-vertex graph $K_L$, where $L=4c$. 
The treewidth of the $N=2cL^2$-vertex Chimera graph comprising an $L\times L$ grid of $K_{c,c}$ unit cells is $cL+1 \sim \mc{O}(\sqrt{N})$ \cite{Choi2}. The treewidth of the $512$-vertex Chimera graph shown in Figure~\ref{fig:chimera} is $33$. Dynamic programming can always find the true ground state of the corresponding Ising model in a time that is exponential in the treewidth, i.e., that scales as $\exp(c\sqrt{N})$.\\

\begin{figure}[t]
\centering
\includegraphics[width=0.9\columnwidth]{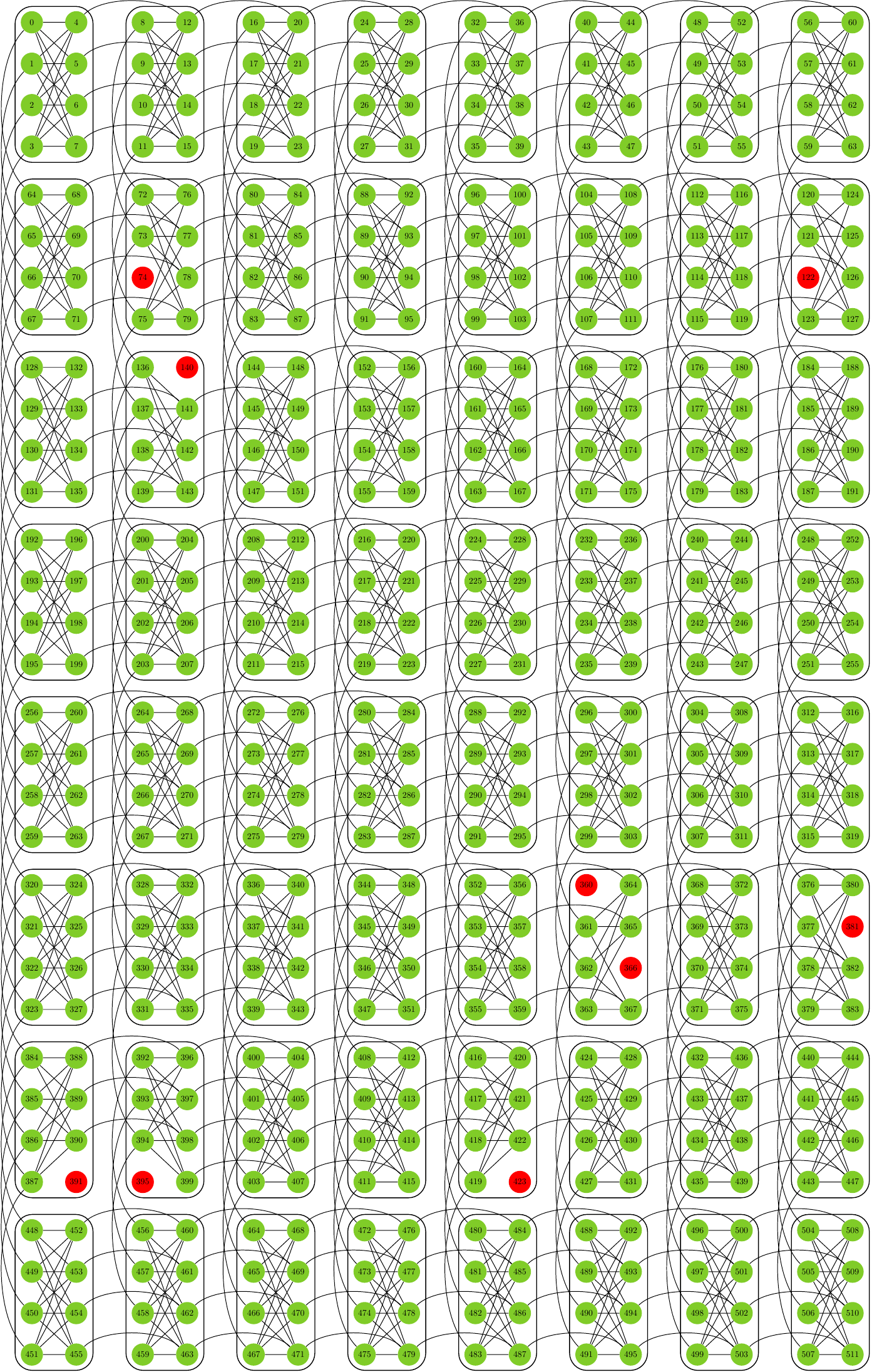}
\caption{\textbf{Qubits and couplers in the D-Wave Two device.} The DW2 ``Vesuvius'' chip consists of an $8\times8$ two-dimensional square lattice of eight-qubit unit cells, with open boundary conditions. The qubits are each denoted by circles, connected by programmable inductive couplers as shown by the lines between the qubits. Of the $512$ qubits of the device located at the University of Southern California used in this work, the $503$ qubits marked in green and the couplers connecting them are functional.}
\label{fig:chimera}
\end{figure}
\begin{figure}
\includegraphics[width=\columnwidth]{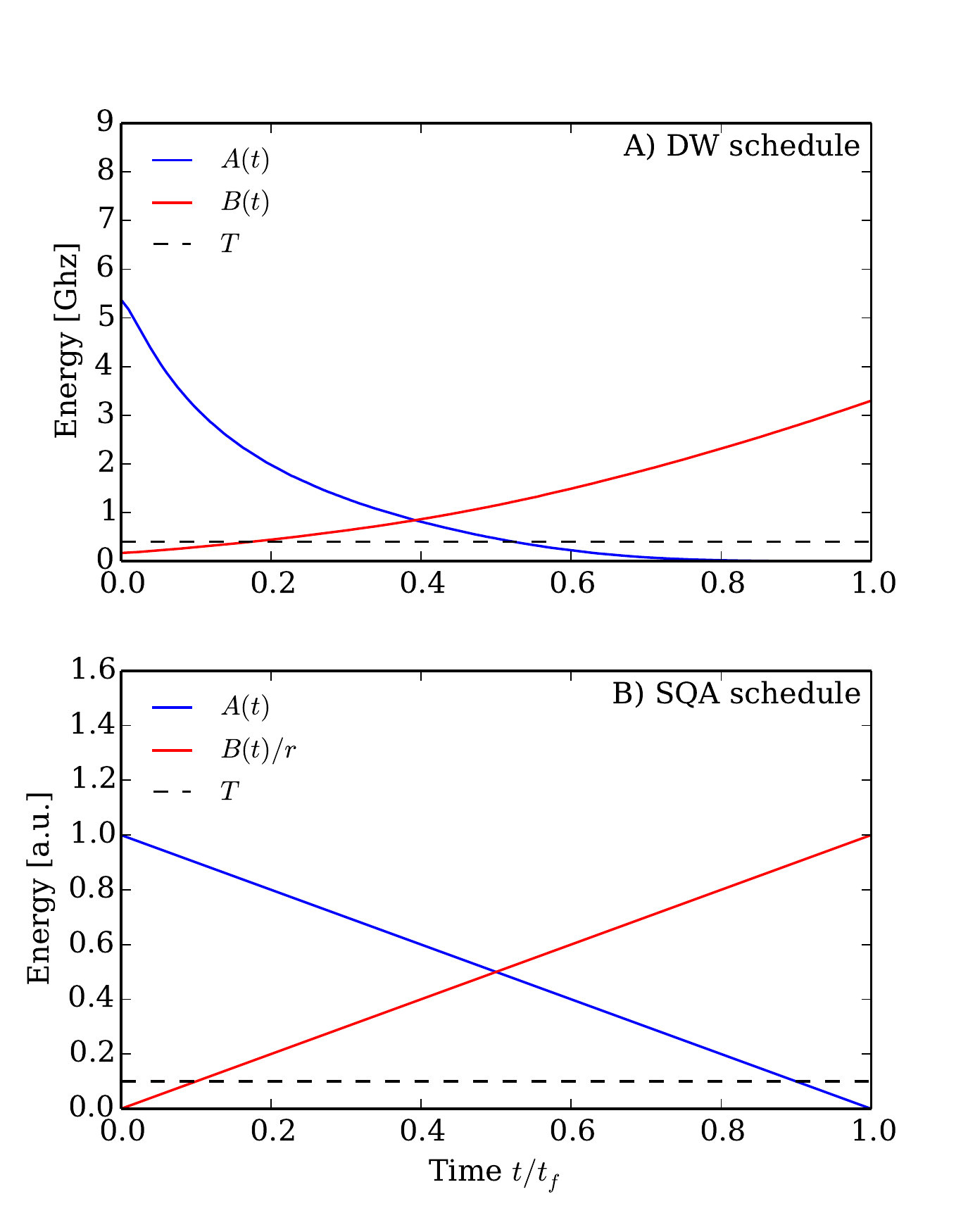}
\caption{{\bf Annealing schedules.} A) The amplitude of the median transverse field $A(t)$  (decreasing, blue) and the longitudinal couplings $B(t)$ (increasing, red) as a function of time. The device temperature of $T=18$mK is indicated by the black horizontal dashed line. B) The linear annealing schedule used in simulated quantum annealing.}
\label{fig:schedule}
\end{figure}
\begin{figure*}[t]
\includegraphics[width=0.95\textwidth]{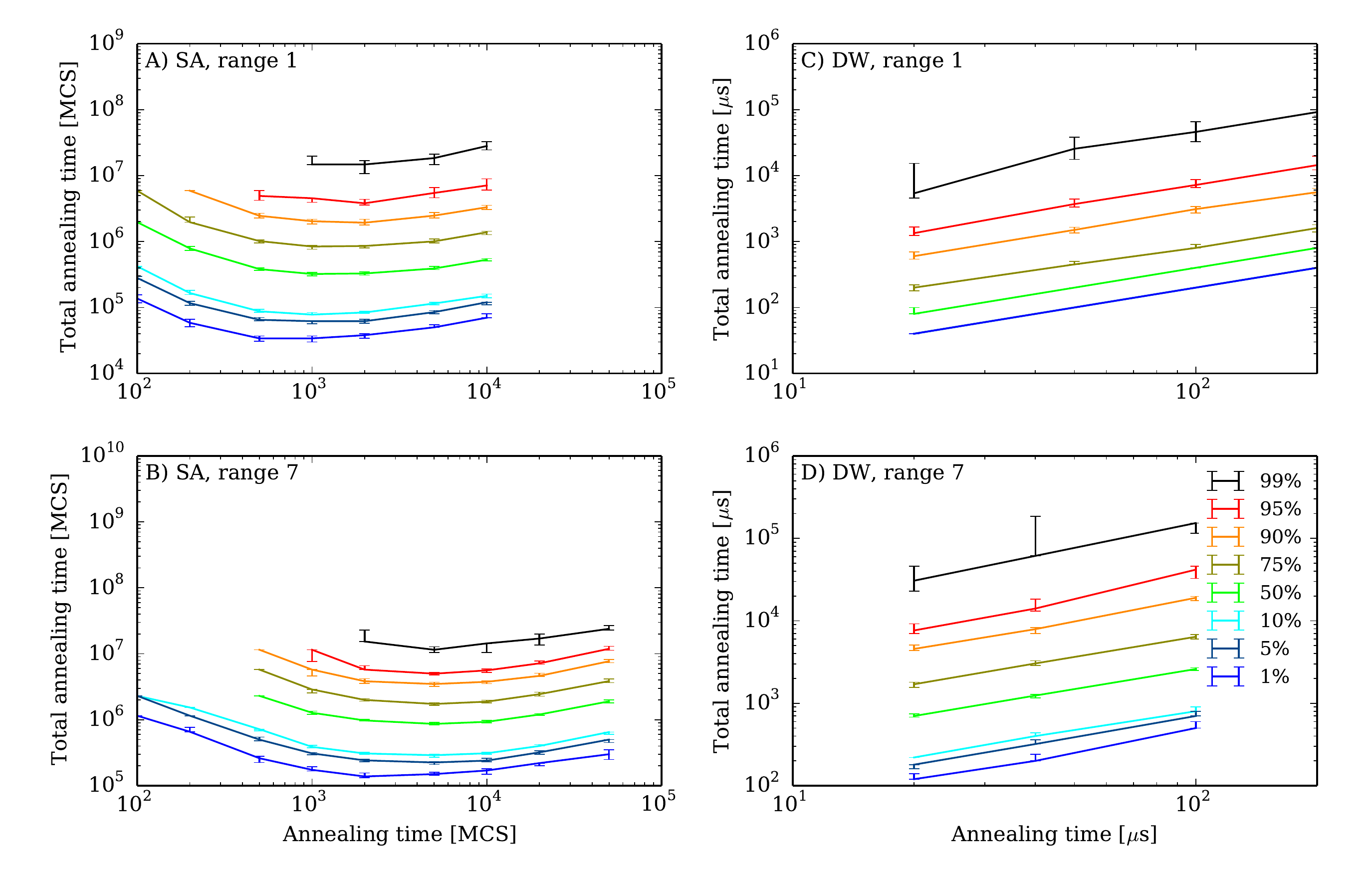}
%\subfigure{\includegraphics[width=0.9\columnwidth]{ssfig03.pdf}}
\vspace{-.5cm}
\caption{{\bf Optimal annealing times for the simulated annealer and for the D-Wave device.} Shown is the total effort $R(t_a)t_a$ as a function of annealing time $t_a$ for various quantiles of problems with $r=1$ and $r=7$ (see Supplementary Information for $r=3$). A) and B) SA, where the minimum of the total effort determines the optimal annealing time $t_a^{\textrm{opt}}$.  C) and D) DW2, where we find a monotonically increasing total effort, meaning that the optimal time $t_a^{\rm opt}$ is always shorter than the minimal annealing time of $20\mu$s.}
\label{fig:satopt}
\end{figure*}

\noindent \textbf{Annealing schedule of the D-Wave Two device.}
Nominally, the DW2 performs annealing by implementing the time-dependent Hamiltonian $H(t) = -A(t) \sum_i \sigma_i^x +B(t) H_{\textrm{Ising}}$, where $t\in [0,t_a]$. However, in reality the transverse field varies somewhat and the actual Hamiltonian realized is described more accurately as 
\begin{equation}
H(t) = -\sum_i A_i(t) \sigma_i^x +B(t) H_{\textrm{Ising}}
\end{equation}
The annealing schedules $A_i(t)$ and $B(t)$ used in the device are shown in figure \ref{fig:schedule}A).
We used the minimal annealing time of $t_a=20\mu s$ provided by the device since this always gave us the shortest total time to solution (see below). The source of this minimal annealing time is engineering restrictions in the DW2. There are four annealing lines, and their synchronization becomes harder for faster annealers. The filtering of the input control lines introduces some additional distortion in the annealing control.

The DW2 is programmed by providing the sets of couplings $J_{ij}$ and local longitudinal fields $h_i$ (which together define a problem instance by specifying $H_{\textrm{Ising}}$), the number of repetitions $R$ of the annealing to be performed, the annealing time $t_a$, and a number of other parameters which were included in our wall-clock results and which are described below.\\

\noindent \textbf{Gauge averaging on the D-Wave device.}
Calibration inaccuracies cause the couplings $J_{ij}$ and $h_i$ that are realized in the DW2 to be slightly different from the intended and programmed values ($\sim 5\%$ variation). These calibration errors can sometimes lead to the ground states of the model realized in the device being different from the perfect model. To overcome these problems it is advantageous to perform annealing on the device with multiple encodings of a problem instance into the couplers of the device \cite{ourpaper}. To realize these different encodings we use a gauge freedom in realizing the Ising spin glass: for each qubit we can freely define which of the two qubits states corresponds to $\sigma_i=+1$ and  $\sigma_i=-1$. More formally this corresponds to a gauge transformation that changes spins $\sigma^z_i\rightarrow  a_i\sigma^z_i$, with $a_i=\pm1$ and the couplings as $J_{ij} \rightarrow a_ia_jJ_{ij}$ and $h_i\rightarrow a_ih_i$. The simulations are invariant under such a gauge transformation, but (due to calibration errors which break the gauge symmetry) the results returned by the DW2 are not.

If the success probability of one annealing run is denoted by $s$, then the probability of failing to find the ground state after $R$ independent repetitions (annealing runs) each having success probability $s$ is $(1-s)^R$, and the total success probability of finding the ground state at least once in $R$ repetitions is
\begin{equation}
 P=1-(1-s)^R .
 \label{eq:oneg}
 \end{equation}
Thus the number of repetitions needed to find the ground state at least once with probability $p$ is found by solving $p=1-(1-s)^R$, i.e., Eq.~\eqref{eq:repetitions}. 

Following \cite{ourpaper}, after splitting these repetitions into $R/G$ repetitions for each of $G$ gauge choices with success probabilities $s_g$, the total success probability becomes
 \begin{equation}
 P^{(G)} = 1-\prod_{g=1}^G(1-s_g)^{R/G}.
 \label{eq:manyq}
 \end{equation}
If we denote by $\overline{s}$ the geometric mean of the success probabilities of the individual gauges
 \begin{equation}
\overline{s} = 1-\prod_{g=1}^G(1-s_g)^{1/G}, 
 \end{equation}
then Eq.~\eqref{eq:manyq} can be written in the same form as Eq.~\eqref{eq:oneg}:
\begin{equation}
 P^{(G)}=1-(1-\overline{s})^R.
 \label{eq:onegav}
 \end{equation}
We thus use the geometric mean $\overline{s}$ in our scaling analysis.\\

\noindent \textbf{Wall-clock and annealing times.}
%\label{sec:wall-clock-annealing-class}
%
We show results mainly for pure annealing times, but also for wall clock times. The pure annealing time for $R$ repetitions is straightforwardly defined as
\begin{equation}
t_{\rm anneal}=Rt_a.
\end{equation}

\begin{table}
\begin{tabular}{|c|c|c|}
\hline 
$N$ & $t_p$ [ms] & $t_r$ [$\mu$s] \\
\hline
8 & $14.7 \pm 0.3$ & $51.0 \pm 0.2$ \\ 
16 & $14.8 \pm 0.3$ & $53.0 \pm 0.2$ \\ 
31 & $14.8 \pm 0.3$ & $57.9 \pm 0.2$ \\ 
47 & $14.9 \pm 0.4$ & $60.6 \pm 0.2$ \\ 
70 & $15.0 \pm 0.4$ & $64.5 \pm 0.2$ \\ 
94 & $15.2 \pm 0.3$ & $68.3 \pm 0.2$ \\ 
126 & $15.6 \pm 0.2$ & $73.1 \pm 0.2$ \\ 
158 & $15.5 \pm 0.2$ & $78.0 \pm 0.2$ \\ 
198 & $15.5 \pm 0.2$ & $80.8  \pm 0.2$ \\ 
238 & $15.7 \pm 0.2$ & $83.5 \pm 0.2$ \\ 
284 & $15.8 \pm 0.2$ & $83.6 \pm 0.1$ \\ 
332 & $16.0 \pm 0.3$ & $87.1 \pm 0.2$ \\ 
385 & $16.6 \pm 1.0$ & $87.1 \pm 0.6$ \\ 
439 & $16.6 \pm 0.1$ & $90.4 \pm 0.1$ \\ 
503 & $16.6 \pm 0.2$ & $90.5 \pm 0.1$ \\ 
\hline 
\end{tabular}
\caption{{\bf Wallclock times on the DW2}. Listed are measured programming times $t_p$ and annealing plus readout times $t_r$ (for a pure annealing time of $20\mu$s) on the DW2 for various problem sizes.}
\label{tab:dw2times}
\end{table}
Wall clock times include the time for programming, cooling, annealing, readout and communication. 
We have performed tests on the DW2 with varying numbers of repetitions $R$ and performed a linear regression analysis to fit the total wall clock time for each problem size to the form $t_p(N)+Rt_r(N)$, where $t_p(N)$ is the total preprocessing time and $t_r(N)$ is the total run time per repetition for an $N$-spin problem. The values of $t_p$ and $t_r$ are summarized in Table \ref{tab:dw2times}. With these numbers we obtain the total wall clock time for $R$ annealing runs split over $G$ gauges (with $R/G$ annealing runs each) as
\begin{equation}
t_{\rm total}(N)=Gt_p(N)+Rt_r(N) .\\
\label{eq:wc}
\end{equation}

\begin{table}
  \centering
  \begin{tabular}{|c|c|c|c|}\hline
 $N\le 238$ & $N=284$, 332 & $N=385$, 439 & $N=503$ \\ \hline
 1000 & 2000 & 5000 & 10000 \\ \hline
  \end{tabular}
  \caption{{\bf Repetitions of annealing runs used on the DW2}. This table summarizes the total number of repetitions used to estimate the success probabilities on the DW2 for various system sizes.}
  \label{tab:reps}
\end{table}

%\section{Classical algorithms}
%\label{sec:classical-programs}

To calculate pure annealing times for the simulated annealer we  determine the total effort in units of  of Monte Carlo updates (attempted spin flips), and then convert to time by dividing by the number of updates that the codes can perform per second \cite{sapaper}. Our classical reference CPU is an $8$-core Intel Xeon E5-2670 CPU, which was introduced around the same time as the DW2. 

To obtain wall-clock times we measure the actual time needed to perform a simulation on the same  Intel Xeon E5-2670 CPU. Since the multi-spin codes perform at $64$ repetitions in parallel, we always make at least $1024$ repetitions when running 16 threads on $8$ cores. This causes the initially flatter scaling in wall-clock times as compared to pure annealing times. The measured initialization time includes all preparations needed for the algorithm to run, and the spin flip rate was computed for the 99\% quantile for 503 qubits. For smaller system sizes or lower quantiles, the spin flip rate is lower since the problems are not hard enough to benefit from parallelization over several cores.\\

\noindent \textbf{Optimal annealing times.}
%\label{sec:scaling-optimality}
As discussed in the main text we need to determine the optimal annealing time $t_a^{\textrm{opt}}$ for every problem size $N$ in order to make meaningful extrapolations of the time to find a solution. To determine $t_a^{\textrm{opt}}$ we perform annealing runs at different annealing times $t_a$, determine the success probabilities $s(t_a)$ of 1000 instances, and from them the required number of repetitions $R(t_a)$ to find the ground state with a probability of 99\%. The total effort $R(t_a)t_a$ diverges for $t_a\rightarrow 0$ and $t_a\rightarrow\infty$ and has a minimum at an optimal annealing time $t_a^{\rm opt}$. The reason is that for short $t_a$ the success probability goes to zero, which leads to a diverging total effort, while for large $t_a$ the time also grows since one always needs to perform at least one annealing run and the total effort is thus bounded from below by $t_a$.

In Figure~\ref{fig:satopt} (left) we plot various quantiles of the total effort $R(t_a)t_a$ for the simulated annealer as a function of $t_a$ to determine the  optimal annealing time $t_a^{\rm opt}$. For the DW2 we find, as shown in Figure~\ref{fig:satopt} (right) that the minimal annealing time of  $20\mu s$  is always longer than the optimal time and we thus always use the device in a suboptimal mode. As a consequence the scaling of time to solution is underestimated, as explained in detail in the main text.\\

\noindent \textbf{Alternative consideration of parallel versus quantum speedup.} In the consideration of how to disentangle parallel and quantum speedup it may seem more natural to assume fixed computational resources of a given device. We will show that this leads to the same scaling as Eq.~\eqref{eq:parallelspeedup1}. 
We might be tempted to define the speedup in this case as
$S(N) = {T_{\textrm{SA}}(N)}/{T_{\textrm{DW}}(N)}$. 
However, 
in this manner only a fraction $N/512$ of the qubits are used while the classical code uses the available CPU fully, independently of problem size. This suboptimal use of the DW2 may again be incorrectly interpreted as speedup. The same issue would appear when comparing a classical analog annealer against a classical simulated annealer.

 As in the discussion of optimal annealing times above, we need to ensure an optimal implementation to correctly assess speedup. For the DW2 (or a similarly constructed classical analog annealer) this means that one should always attempt to make use of the entire device: we should perform as many annealing runs in parallel as possible.
Let us denote the machine size by $M$ (e.g., $M=512$ in the DW2 case). With this we define a new, optimized, annealing time
\begin{equation}
T^{\rm opt}_{\textrm{DW}}(N) = T_{\textrm{DW}}(N) \frac{1}{\lfloor M/N\rfloor} ,
\label{eq:T^opt_DW}
\end{equation}
and the correct speedup in our case is then
\begin{equation}
S(N) = \frac{T_{\textrm{SA}}(N)}{T^{\rm opt}_{\textrm{DW}}(N)} =  \frac{T_{\textrm{SA}}(N)}{T_{\textrm{DW}}(N)} \left\lfloor\frac{M}{N}\right\rfloor.
\label{eq:parallelspeedup2}
\end{equation}

Omitting the floor function ($\lfloor \; \rfloor$), which only gives subdominant corrections in the limit $M\rightarrow\infty$ we recover Eq.~(\ref{eq:parallelspeedup1}).

%\noindent \textbf{Derivation of Eq.~\eqref{eq:T^opt_DW} from the scaling of the annealing time.}
The conclusion that the speedup function includes a factor proportional to $1/N$ is validated from yet another perspective, that focuses on the annealing time. Instead of embedding $C\equiv \lfloor{M}/{N}\rfloor$ different instances in parallel, we can embed $C$ replicas of a given instance.
Each replica $r$ (where $r\in\{1,\dots,C\}$) results in a guess $E_{r,i}$ of the ground state energy for the $i$th run, and we can take $E_i = \min_r E_{r,i}$ as the proposed solution for that run. 
If the replicas are independent and each has equal probability $s$ of finding the ground state, then 
using $C$ replicas the probability that at least one will find the ground state is $s' = 1-(1-s)^C$, which is also the probability that $E_i$ is the ground state energy for the $i$th run. Repeating the argument leading to Eq.~\eqref{eq:repetitions}, the number of repetitions required to find the ground state at least once with probability $p$ is then:
\begin{equation}
  R' = \left\lceil
    \frac{\log (1-p)}{\log(1 - s')}\right\rceil 
    = \left\lceil
    \frac{\log (1-p)}{C\log(1 - s)}\right\rceil
    =\frac{R}{C} .
\end{equation}
Focusing on the pure annealing time we have $T^{\rm opt}_{\textrm{DW}}(N) = t_a R'$ and $T_{\textrm{DW}}(N) = t_a R$, which yields Eq.~\eqref{eq:T^opt_DW}.}\\

%\input{methods11}
%%%%%%%%%

%%%%%%%%%

{\small \noindent \textbf{Acknowledgements}\\ 
\noindent We thank N. Allen, M. Amin, E. Farhi, M. Mohseni, H. Neven, and C. McGeoch for useful discussions and comments. We are grateful to I. Zintchenko for providing  the  {\tt an\_ss\_ge\_nf\_bp} simulated annealing code before publication of the code with Ref.~\cite{sapaper}.
This project was supported by the Swiss National Science Foundation through the National Competence Center in Research NCCR QSIT, the ARO MURI Grant No. W911NF-11-1-0268, ARO grant number W911NF-12-1-0523, the Lockheed Martin Corporation and Microsoft Research. We acknowledge hospitality of the Aspen Center for Physics, supported by NSF grant PHY-1066293.}
\\

%\input{ack11}
%%%%%%%%%

%%%%%%%%%
%{\small 
%\noindent \textbf{Author contributions:} All authors contributed to the planning of the tests, the discussions presented in the paper, and data analysis. Data on the DW2 was taken by ZW, JJ and SB. Simulation codes were written by TFR and SVI, and simulations were performed by TFR, SVI and DW. TFR and SB wrote the exact classical solvers. MT and DL wrote the manuscript with input from the other authors.\\

%\noindent \textbf{Additional information}\\
%\noindent Supplementary Information accompanies this paper.\\\\

%\noindent The authors declare no competing financial interests.
}
%\input{cont11}
%%%%%%%%%

\bibliographystyle{naturemag}
\bibliography{speedup}

\begin{thebibliography}{10}
\bibitem{P-NP-comment}
\bibinfo{note}{{Such a proof seems unlikely to be found any time soon since it
  would imply that factoring is not in the complexity class P (polynomial) and
  thus P and NP (nondeterministic polynomial) are distinct, solving the
  long-standing P versus NP question.}}
\bibitem{community-comment}
\bibinfo{note}{{For example, it may be the case, though it seems unlikely, that
  a classified polynomial-time factoring algorithm is available to parts of the
  intelligence community.}}
\bibitem{opt-comment}
\bibinfo{note}{{While one can in principle look for an optimal annealing time
  for each individual problem instance, we instead determine an averaged
  optimal annealing time $t_a^{\rm opt}(N)$ for each problem size $N$ by
  annealing many instances at various annealing times $t_a$, and then use these
  for all future problems of that size.}}
\bibitem{hybrid-comment}
\bibinfo{note}{{The fact that some problems are solved faster on classical
  hardware and some on the DW2 raises the possibility of a hybrid approach that
  benefits by solving a problem instance using both, and then selecting the
  best solution found.}}
\bibitem{MAX2SAT-comment}
\bibinfo{note}{{For example, it is possible that a randomized benchmark with a
  tunable ``hardness knob", such as the clause density in the MAX-2SAT problem
  \cite{MAX2SAT}, will allow for a more fine-tuned exploration of the
  performance potential of a quantum annealer than a purely randomized
  benchmark such as we have used here. It was also proposed that embedding 3D
  spin-glass problems into the Chimera graph, with a finite critical
  temperature, might yield problem classes that are better suited as benchmarks
  for quantum annealing \cite{katzgraber}.}}
\end{thebibliography}

\end{document}